\documentclass[aps,prl,twocolumn,superscriptaddress]{revtex4}

\usepackage{amssymb}
\usepackage{graphicx}

\newcommand{\ket}{\rangle}
\newcommand{\bra}{\langle}

\begin{document}
\bibliographystyle{apsrev}

\title{A numerical canonical transformation approach to quantum many body
problems}
\author{ Steven R.\ White}
\affiliation{ 
Department of Physics and Astronomy,
University of California,
Irvine, CA 92697
}
\date{\today}
\begin{abstract}
\noindent 

We present a new approach for numerical solutions of {\it ab
initio} quantum chemistry systems.
The main idea of the approach, which
we call {\it canonical diagonalization}, is to diagonalize
directly the second quantized Hamiltonian by a sequence of
numerical canonical transformations.
\end{abstract}
\maketitle


\section{Introduction}
The vast majority of current methods for numerically solving 
quantum many body systems
fall into a few broad categories. One type of approach involves
producing approximate representations of wavefunctions. Examples
of this approach include configuration interaction, coupled
cluster methods\cite{cc}, and the density matrix renormalization group 
(DMRG)\cite{dmrg,WhiteMartin,mitrushenkov}.
A second type of approach is based on perturbation theory. Some
versions of perturbation theory are closely related to
wavefunction approaches, while others, for example utilizing
finite temperature imaginary time Green's functions, are more
closely related to path integrals.
A third type of approach is quantum Monte Carlo, which also may
involve representations of wavefunctions (e.g. Green's function Monte
Carlo) or of path integrals (e.g. determinantal/auxiliary field
methods).

Much less explored than these approaches are methods based
on similarity or unitary transformations of the Hamiltonian $H$.
In these approaches the primary focus is on $H$
and transformed versions of it; wavefunctions play a much
more minor role.
In this paper we develop such an approach in the context
of {\it ab initio} quantum chemistry calculations in a
finite basis. This approach is based on 
unitary canonical transformations (CTs) of $H$ written in 
second-quantized operator form. Such transformations have been
used in analytical work for very long time\cite{vanvleck}. A
well known example in condensed matter physics is the
Schrieffer-Wolff transformation of the Anderson model into the
Kondo model\cite{schrieffer}; another is the well known mapping
of the Hubbard model in the large $U/t$ limit into the $t$-$J$
model.  Often these transformations are performed once, and
relate one model system to another, simpler system, with fewer
degrees of freedom, in an approximate way.
In some cases
special CTs can produce exact solutions for certain model systems.

Recently, substantial progress has been made by the development 
of continuous unitary transformations, in which a set of
differential equations is solved to perform the
CT.\cite{wegner,glazek} This method, which was developed
independently by Wegner and by Glazek and Wilson, is known by the names
``flow equation method" \cite{wegner} and
``similarity renormalization" \cite{glazek}. A key advantage of
this approach is that one does not need to know in advance the 
transformation operator to be used; it is determined implicitly by the
solution of the differential equations. Another advantage is
that once the differential equations are set up, there is no
operator algebra to be performed in the course of the
numerical solution of the differential equations.

This approach can be performed in a semianalytic context, where
typically one is deriving one model from another. For example,
an improved treatment of the Schrieffer-Wolff transformation of
the Anderson model has been performed. \cite{kehrein} It can
also be used to obtain ground state energies and dispersion
relations (i.e. excitation energies) 
for infinite lattice systems. \cite{knetter} 

Here we extend and develop the CT approach in a quantum chemical
context. We first introduce a new version of the CT approach which
is closely related to the Jacobi method of diagonalizing
matrices.  In this case, rather than solving a differential
equation, one performs a sequence CTs, each involving the
smallest possible number of operators. This approach is
designed to solve a finite quantum system, namely a molecule or
cluster in a standard quantum chemical basis, as an alternative to
other standard approaches, such as configuration interaction 
or coupled cluster methods.  We demonstrate that our method can
determine ground states 
in {\it ab initio} chemical systems with excellent precision,
utilizing tests on a water molecule for which exact results
are available.
We expect similar performance for low lying excited states.
We then present a version of the ``flow equation" approach for
use in the {\it ab initio} context. Utilizing ideas developed
in our Jacobi method, we present a new version of the
differential equations which make their numerical solution
particularly efficient. We demonstrate that this method also
works very well for the water molecule.

Perhaps the most important aspect
of our approaches is the ability to remove both low
energy (i.e. core) and high energy virtual orbitals from the problem,
leaving a system with a small number of ``active" orbitals.
This approach is especially useful for
``strongly correlated" systems, i.e. those with open shells,
breaking bonds, etc., where a single reference approach fails.
In these systems, the strong correlation is generally confined
to a relatively small number of orbitals. In wavefunction-based 
approaches, it can be awkward and expensive to deal with the strongly
correlated part of the problem with a powerful method (e.g.
full diagonalization of the active space) and the simple
high-energy part of the problem with another, simpler method.
By working with the Hamiltonian directly, one can separate
the two parts of the problem simply and efficiently. First one
transforms the whole Hamiltonian into a form in which the high
and low energy orbitals are either completely unoccupied or complete
occupied (for core orbitals). Then these orbitals
can be thrown out of the problem completely, 
leaving a smaller system of partially occupied orbitals which 
can then be solved with any of a
variety of nonperturbative techniques. Here we demonstrate this
powerful hybrid approach on a stretched water molecule, solving the
smaller systems with DMRG.

Another advantage in working with the Hamiltonian directly is
in obtaining excitations. In transforming the Hamiltonian, we
make progress in solving for excited states at the same time
as we obtain the ground state. Essentially the same approach
is used to obtain both ground and excited states. Although we have
not performed tests yet for excited states, CT approaches in
other contexts have obtained excellent results. \cite{knetter}

We will call the set of methods we present here 
canonical diagonalization (CD).
The term diagonalization indicates our intent to solve
the system fully, rather than just transforming to a simpler
model. The transformations involved are both unitary and
canonical (which is defined in the next section); we choose
only the term canonical to emphasize that the method works
in the space of second quantized operators. We further
distinguish the Jacobi CD method (JCD) and the flow equation
CD method (FECD).
All CD approaches share the feature that the
object that one is manipulating is the second quantized
Hamiltonian, as a collection of abstract operator terms with
specific numerical coefficients.

These approaches seem particularly suited to {\it ab initio} quantum 
chemistry, which are characterized by a very
general quantum Hamiltonian, containing almost all possible one
and two-electron terms. The CTs generate
additional terms involving one, two, three, and more particles. 
Since general one and two particle terms are already present in
the Hamiltonian, no extra inconvenience arises from these terms.
Three and more particle terms are more inconvenient, but most such
terms can be neglected, to an excellent approximation. CD is
size-consistent, and many-particle terms which may be left out
involve the simultaneous interaction
of three or more (dressed) electrons, so that neglecting them is analogous
to neglecting connected clusters involving triple and higher
excitations in coupled cluster methods.  
Note that canonical transformations also appear in 
the theory of the coupled cluster method\cite{nooijen}, although
the method remains largely a wavefunction approach.
Note also that although one does not need to write any wavefunctions
explicitly, CD in its simpler forms
implicitly expresses the ground state using the exponential
of an operator acting on a reference state.  Further links
to coupled cluster methods are made in the Discussion Section.

CD fits naturally into a renormalization group (RG) framework.
First, one can remove (``integrate out'') higher energy
orbitals, one at a time if one wishes, leaving a system where
the effects of the 
removed orbitals are incorporated into an effective Hamiltonian for
the remaining orbitals. Thus each step resembles a
transformation in a typical RG calculation in statistical
mechanics, although unlike in that case one cannot continue
indefinitely and there are no fixed points. 
Second, even if one is not integrating out orbitals,
the transformation of the Hamiltonian, like in RG methods,
occurs in a sequence of steps, with
truncation of higher order terms occuring at each step.
Third, the differential ``flow" equation form of CD,
in which a time-like
variable controls the evolution of the Hamiltonian operator towards
a more diagonal form, closely matches Wilson's
original conception of the RG approach\cite{wilsonrg}.

CD is a natural complement to DMRG, and this was a principle motivation
in developing it. When applied to quantum chemistry problems,
DMRG does very well in describing non-dynamical correlations
(strong correlations associated with partially occupied orbitals),
but it is inefficient in describing dynamical correlations, since
it describes high energy virtual orbitals on the
same footing as partially occupied strongly interacting
orbitals\cite{chan}. CD has complementary behavior. It can
be used to remove the nearly unoccupied orbitals, leaving
a smaller Hamiltonian involving strongly interacting orbitals 
for DMRG to solve.

\section{Jacobi CD}
We begin with the Jacobi CD approach.
In the ordinary Jacobi method for diagonalizing matrices, 
one applies a large number of
unitary transformations to a Hermitian matrix to bring it into
diagonal form\cite{numrecipe}. A unitary transformation gives
a new matrix which has the same eigenvalues as the old.
In the Jacobi method,
each unitary transformations consists of a
rotation of two rows and columns to zero out a single
off-diagonal element $H_{ij}$. The part of the unitary transformation
matrix $\exp(A)$ corresponding to rows and columns $i$, $j$ is
\begin{equation}
\exp \left( \begin{array}{ccc}
0 &  \theta \\
-\theta & 0 
\end{array} \right)  =
\left( \begin{array}{ccc}
\cos\theta &  \sin \theta  \\
-\sin\theta & \cos \theta  
\end{array} \right)
\end{equation}
The angle $\theta$ which removes the term $H_{ij}$ is given by
\begin{equation}
\theta = \frac{1}{2} \tan^{-1}[2 H_{ij} /(E_i-E_j)]
\end{equation}
where $E_i = H_{ii}$ 
and the transformation is applied as
\begin{equation}
\exp(A) H \exp(-A).
\end{equation}
Any unitary transformation can be written as an exponential of
an antiHermitian matrix $A$\cite{wagner}. For real symmetric matrices and 
operators $H$ and for
what follows, it suffices to use real antisymmetric matrices
and operators $A$.
In the Jacobi method, one traverses
the matrix repeatedly, rotating away off-diagonal elements, 
starting with the largest off-diagonal elements for efficiency.

In Jacobi CD we construct unitary transformations to successively remove
off-diagonal terms of a second quantized Hamiltonian.
We consider a quantum chemical system in a Hartree
Fock basis, with Hamiltonian
\begin{equation}
H = \sum_{ij\sigma} T_{ij} c^\dagger_{i\sigma}c_{j\sigma}
+ \frac{1}{2} \sum_{ijkl\sigma\sigma'} V_{ijkl}
c^\dagger_{i\sigma}c^\dagger_{j\sigma'}c_{k\sigma'}c_{l\sigma} .
\label{Ham}
\end{equation}
Here $T_{ij}$ contains the electron kinetic energy and the
Coulomb interaction between the electrons and the nuclei, while
$V_{ijkl}$ describes the electron-electron Coulomb interaction.
Indices such as $i$ denote spatial Hartree-Fock orbitals, and $\sigma$
is a spin index. Later on we shall sometimes use $i$ to denote
a spin-orbital, in which case the context should make the usage
self-evident.
We use a computer representation of $H$ as a sum of abstract
operator terms, each with a coefficient and a product of $c$ and $c^\dagger$
operators involving various orbitals. In our program a $c$ or
$c^\dagger$ operator is described by a single byte, which
encodes the orbital index involved, the spin, and whether it is
$c$ or $c^\dagger$. This implementation is thus limited to
systems with at most 64 orbitals. A complete operator term is stored as an
array of such bytes plus a floating point coefficient.  
We developed a set of
C++ routines to take products and commutators of such operators, 
putting the result
in normal ordered form using the anticommutation relations.
Having these operations be reasonably efficient
is crucial to the method. This approach, using formal operator terms 
to describe $H$, rather than specific matrix expressions for 
terms of various orders, is both simple and general. In the future, 
in order to implement specific approximations within CD,
more efficient code could be produced by deriving and
implementing the relevant matrix expressions.

An off-diagonal term which can be rotated away is simply any
term which is distinct from its Hermitian conjugate.
Self-adjoint terms constitute the diagonal elements. 
They can always be written as products of density operators
$n_{i\sigma} = c^\dagger_{i\sigma} c_{i\sigma}$.
Consider,
as a specific example, the term 
\begin{equation}
V_\alpha = a \bar V_\alpha = 
a c^\dagger_{i\uparrow}c^\dagger_{j\downarrow}
c_{k\downarrow}c_{l\uparrow} ,
\label{Valpha}
\end{equation}
where $a$ is a numerical coefficient.  Let 
\begin{equation}
A = \theta (\bar V_\alpha - \bar V_\alpha^\dagger)
\label{Adef}
\end{equation}
Given the proper choice of $\theta$, the transformation 
$H \to \exp(A) H \exp(-A)$
will remove $V_\alpha + V_\alpha^\dagger$ from $H$, introducing
other terms instead. 
We will consider the choice of $\theta$ momentarily.
These additional terms in general have smaller coefficients than
$V_\alpha$, making $H$ more diagonal. If one continued the
process indefinitely, one would have a ``classical" Hamiltonian
where every term was diagonal. 
In this case any Slater determinant 
$c^\dagger_i c^\dagger_j \ldots |0\ket$ is an eigenstate and
all eigenvalues can be read off essentially by inspection. 
After the particle-hole
transformation of the occupied orbitals described below,
normally the ground state energy is simply the constant term in
$H$. 

This unitary transformation is a canonical transformation, in
that when applied to the operators $c$ and $c^\dagger$, the
anticommutation relations are preserved, e.g. 
\begin{equation}
\{e^A c_i e^{-A},e^A c^\dagger_j e^{-A}\}
= e^A \{c_i , c^\dagger_j\} e^{-A}
= \{c_i , c^\dagger_j\} .
\label{comrels}
\end{equation}
One can view a CT as a complicated
many-particle change of basis. In that sense CD is similar
to the DMRG method, the biggest difference being that DMRG
works in a wavefunction basis. In the Discussion section,
we comment more on applying the CTs to the $c$ operators.

In order to carry out the transformations, it is convenient to
use the well-known formula
\begin{eqnarray}
e^A H e^{-A} &=& H + [A,H] + \frac{1}{2!}[A,[A,H]] \nonumber\\&+&
\frac{1}{3!}[A,[A,[A,H]]] + ...
\label{expandA}
\end{eqnarray}
A commutator between an $N_1$-particle term and an
$N_2$-particle term gives up to $(N_1+N_2-1)$-particle
terms. Hence, if $A$ is a one particle term, it does not
generate any higher order terms. Indeed, rotations by
one-particle $A$'s correspond to single particle changes of
basis. CD using one-particle $A$'s can be used to perform a 
Hartee-Fock calculation,
if the initial orbitals are not the HF orbitals.
To evaluate Eq. (\ref{expandA}),
we generally treat each term in $H$ separately. For
most terms $V_\beta$ the transformation has no effect, since $[A,V_\beta] = 0$.
For relevant terms the expansion can be carried out order by
order until all terms of a given order are neglible (i.e. their
coefficients are very small). At each
order the terms should be put in normal ordered form.
The number
of distinct terms generated from a single term $V_\beta$ is a
modest finite number, since a particular  $c_i$  or
$c^\dagger_i$ can appear at most once in a term in its normal
ordered form.

For several reasons it is convenient to perform a particle hole
transformation on the occupied orbitals. Thus we define
$d_{i\sigma} = c_{i\sigma}$ if orbital $i$ is unoccupied in the HF
state, and $d_{i\sigma} = c^\dagger_{i\sigma}$ if it is occupied.
The $d$'s have identical anticommutation relations to the $c$'s.
After this transformation, anticommutation relations are used to put
the terms in normal ordered form, putting
$d$ to the right of $d^\dagger$. The resulting anticommutators generate
new lower order terms, after which the HF energy appears as a constant term
in $H$. In terms of the $d$ operators, the HF state is the
vacuum $|0\rangle$. There are no off-diagonal single particle terms such as
$d^\dagger_i d_j$. There are terms which appear to violate
particle conservation, such as $d^\dagger_i d^\dagger_j d^\dagger_k d^\dagger_l$,
but which in fact do not in terms of real particles.
As one performs the canonical transformations, the vacuum
state approaches the exact ground state.

We now consider the choice of $\theta$. An operator term
$V_\alpha$ connects an exponentially large number of pairs of states $l$,$r$ 
together, $\langle l | V_\alpha | r \rangle \ne 0$. However, one
of these pairs of states can be considered the most important,
namely, the one closest to the HF vacuum. 
This pair has the
fewest number of $d^\dagger$'s operating on $|0\rangle$ which
generate states not destroyed by $V_\alpha$.
As a specific example, let $V_\alpha = 0.1\ 
d^\dagger_i d_j d_k d_m$. Then the most important pair of states
is
\begin{eqnarray}
| r \rangle &=& d^\dagger_j d^\dagger_k d^\dagger_m |0\rangle\nonumber\\
| l \rangle &=& d^\dagger_i |0\rangle .
\end{eqnarray}
We will call these states simply the left and right states of
$V_\alpha$.
Other pairs of states, considered to be less important, have
additional $d^\dagger$'s, which do not appear in $V_\alpha$,
applied to both $|l\rangle$ and $|r\rangle$. For example, one could
take the pair $d^\dagger_n |l\rangle$ and $d^\dagger_n|r\rangle$.
Define for a term $V_\alpha$ 
the left energy $E_l = \langle l | H | l \rangle$ and
similarly the right energy $E_r = \langle r | H | r \rangle$.
We now choose to use the Jacobi formula for $\theta$ to attempt
to eliminate the off-diagonal term in the Hamiltonian connecting
$l$ and $r$. Thus, if $a$ is the coefficient of
$V_\alpha$, we choose
\begin{equation}
\theta = \frac{1}{2} \tan^{-1}[2 a /(E_l-E_r)] .
\end{equation}
This does not eliminate $V_\alpha$ exactly, since the operator
$A$ connects many different states, not just $l$ and $r$, so it
is not a $2\times2$ transformation, as it would be for a matrix.
Nevertheless, we find that this choice generally works well, typically 
reducing the size of the coefficient of $V_\alpha$ by a few
orders of magnitude. Note that the degenerate case $E_l=E_r$ is
nonsingular, generating an angle of $\pm \pi/4$ (either angle
can be chosen). Such a large transformation angle should be
avoided if possible, however, since it generates high order
terms in the transformation of $H$.

A more common choice in analytic work using CTs is to choose to
eliminate $V_\alpha$ to first order in the expansion Eq.
(\ref{expandA}), namely choosing $\theta$ to set the coefficient of
$\bar V_\alpha$ in  $[A,H^D]$ to $-a$, 
where $H^D$ is the diagonal part of the Hamiltonian. This is
closely related to our approach: note that 
$\langle l | [\bar V_\alpha,H^D] | r \rangle$ is the coefficient of
$\bar V_\alpha$ in  $[\bar V_\alpha,H^D]$. However,
\begin{equation}
\langle l | [\bar V_\alpha,H^D] | r \rangle = (E_r-E_l)
\langle l | \bar V_\alpha | r \rangle = E_r-E_l
\end{equation}
so that this choice gives  $\theta (E_r-E_l) = -a$. This agrees
with our choice to lowest order, but it is not well behaved if
$E_r \approx E_l$.

One need not eliminate all off-diagonal terms in $H$. If one is
interested in only the ground state, then one needs to eliminate
all terms connecting that state to other states. More
specifically, suppose the initial HF state $|0\ket$ has
substantial overlap with the ground state. Only states which
produce a nonzero result when acting on the vacuum need be
removed, namely, only terms such as 
$d^\dagger_j d^\dagger_k d^\dagger_m d^\dagger_n$ or
$d^\dagger_j d^\dagger_k$, and their Hermitian conjugates
(as well as similar multiparticle terms). Since all terms still
satisfy particle conservation, in each of the two-particle terms 
two of the orbital indices 
$jkmn$ must correspond to occupied states, and two to unoccupied orbitals.
Once all such terms are eliminated, then the state $|0\ket$ is
the ground state, and the ground state energy is the constant
term in the Hamiltonian.

In all but the smallest systems, some of the terms formed from
the CTs must be discarded according to some criterion. The
simplest criterion is to neglect all terms involving three or
more particles, i.e. six or more $d$ operators. Our test
calculations on the water molecule suggest that this
is a very accurate approach for systems well described by a
single reference state. Other possible criteria include keeping
all terms whose coefficients are larger than some cutoff; keeping 
all one and two particle terms and all three particle terms
larger than a cutoff, etc. More sophisticated criteria are
possible also, such as trying to estimate the contribution of
each term using pertubation theory, and discarding terms whose
contribution is below a cutoff. Here, we perform some 
test calculations according to simple cutoff criteria. In the
future, we hope the criteria can be optimized.

In order to preserve symmetries, such as spin symmetry,
one can rotate sets of terms which are related by a symmetry
transformation in one step. For example, in what follows, for
each term, we check to see if it is distinct from the term
coming from flipping all of its spin indices. If it is distinct,
both are rotated together with the same rotation angle. The
rotation angle for both is chosen as the angle to rotate one of the
terms separately. This procedure preserves spin symmetry
exactly.

One must decide in which order to go through the terms
in performing the CTs. Since each CT alters the coefficients
of many other other
terms in H, it makes sense to start with the largest first.
One approach would be to find the
term with the largest magnitude coefficient at each step.
Another would be to choose the largest rotation angle. 
However, searching for the largest term at each step would be inefficient.
Therefore, we have chosen the following method: a cutoff
angle is chosen, and all terms with angles greater in magnitude
than this cutoff are treated in a sweep through the terms, 
in a predetermined but arbitrary order.
Then, the cutoff angle is reduced
by a constant factor, and the procedure is repeated. Here, we
started with a cutoff of 0.15 and reduced it using a factor of 0.6.
(In some passes, particularly the initial one, there may be no
CTs performed.)
In Table I we show results for a 25 orbital DZP basis water
molecule, for which full CI results are
available\cite{bauschlicher}. Because of some arbitrary choices in
the ordering of the CTs, which unfortunately can affect the results slightly,
the results here would be difficult to reproduce precisely by
an independently written program. (The differential equation
method discussed below does not suffer this problem.) Despite
this, one can easily evaluate the potential of the method from
our results. One can see that we obtain an accuracy of several
millihartrees even when truncating all three or more particle
terms. If one keeps also some three particle terms, one can
obtain accuracy to fractions of a millihartree. This sort of
accuracy is comparable to coupled cluster methods.

\begin{table}
\caption{
Results from the Jacobi CD method applied to the water molecule
in a 25 orbital basis. The 1s O core HF orbital has been
frozen. The exact energy of the system in this basis is -76.256624 
Hartrees.
In all cases, all two particle terms have been retained.
$\varepsilon_3$ and $\varepsilon_4$ are the cutoffs for
retaining three and four particle terms, and $N_3$ and $N_4$
are the corresponding maximum number of such terms in $H$
during the diagonalization. $\Delta E$ is the error in the
energy, $E-E_{\rm exact}$.
}\label{tableone}
\begin{tabular}{l l l l l l}
$\varepsilon_3$ & $\varepsilon_4$ & $\Delta E$ & $N_3$ & $N_4$ \\
\hline
$\infty$ & $\infty$ & 0.0041 & 0 & 0\\
0.01 & $\infty$ & 0.0041 & 464 & 0\\
0.001 & $\infty$ & 0.0019 & $1.1\times10^5$ & 0\\
0.0005 & $\infty$ & 0.0011 & $3.6\times10^5$ & 0\\
0.0001 & $\infty$ & -0.0001 & $2.9\times10^6$ & 0\\
0.0001 & 0.0001 & -0.0001 & $2.9\times10^6$ & $2.0\times10^5$
\end{tabular}
\end{table}

It is not necessary to perform every CT in this procedure.
At the end of every sweep, we perform a calculation of the
energy using second order perturbation theory for the
current Hamiltonian. The calculation time for this procedure 
scales only as the number of terms in H, so there is a neglible
impact on the overall computation time.
As the largest angle terms are eliminated,
the perturbation theory result becomes more and more accurate.
Even just a few rotations of the largest terms can make
perturbation theory much more well-behaved.
One can stop the procedure when the perturbation result
is well converged, which typically happens long before all the
chosen terms are removed. In Fig. 1 we show the results for
this procedure. One can see that the perturbation result
converges much more quickly than the constant term in the
energy. One might well stop after about 300 Jacobi steps;
in this particular example this number is of order $N^2$, 
where $N$ is the number of orbitals.

\begin{figure}[tb]
\includegraphics*[height=7cm]{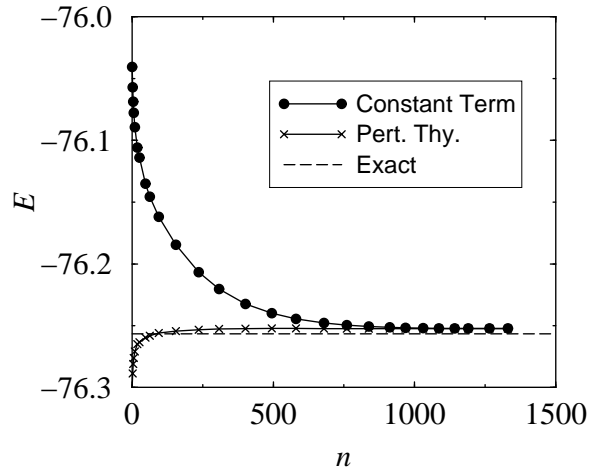}
\caption{
Energy for the water molecule of Table 1 as a function of the
number of Jacobi rotations performed $n$. At each sweep the
constant term of $H$ is shown, as well as the current
result from second order perturbation theory.
The initial value of the constant term is the HF energy;
the initial value of the perturbation theory is what one
would get from it without doing CD.
}
\label{enfigjac}
\end{figure}

In Table II we show similar results for a water molecule
whose bonds have been stretched by a factor of two. This system
is not well described by a single reference state: in the
full CI calculations of Olsen, et. al.\cite{olsen}, in a different but
similar basis, the weight of the HF determinant in the full CI
wavefunction was 0.589, versus 0.941 for the unstretched
molecule. Here, we find that CD is unstable if only two-particle
terms are kept. One finds that repeated Jacobi diagonalization
steps reduce the energy without bound. CD is exact if no
truncations are made, so this is an artifact of the truncation
of three and more particle terms. However, keeping even a large
number of three particle terms does not result in a particularly
accurate calculation.

\begin{table}
\caption{
Same as for Table I, but for the water molecule with bond
stretched by a factor of two. The exact result is -75.95227.
}\label{tabletwo}
\begin{tabular}{l l l l l l}
$\varepsilon_3$ & $\varepsilon_4$ & $\Delta E$ & $N_3$ & $N_4$ \\
\hline
$\infty$ & $\infty$ & $-\infty$ & 0 & 0\\
0.01 & $\infty$ & 0.007 & $1.6\times10^4$ & 0\\
0.001 & $\infty$ & 0.015 & $1.9\times10^5$ & 0\\
0.0005 & $\infty$ & 0.015 & $3.8\times10^5$ & 0\\
0.01 & 0.01 & 0.032 & $1.3\times10^4$ & $3.9\times10^3$
\end{tabular}
\end{table}

For this system, examination of the occupancies 
of the HF orbitals in the exact
ground state (which we have computed with high accuracy with
DMRG) reveals that there
are four spatial orbitals with occupancies far from 0 or 2;
specifically, they have occupancies of 1.58, 1.52, 0.46, and
0.4. The rest have occupancies less than 0.03 or more than 1.97.
In the case of the unstretched water molecule, occupancies are
all within 0.05 of 0 or 2.
The results for occupancies of natural orbitals are very similar\cite{olsen}.
The contribution to the energy of a Hamiltonian term $A$,
$\bra \psi | A | \psi \ket$, can be expressed as a Green's function or density
matrix element. In the case of
Hamiltonian terms made out of operators involving only nearly filled 
or unfilled orbitals, the behavior of the Green's function is well
understood, and the magnitude falls rapidly as one considers
terms involving more particles. 
For partially occupied orbitals there is no
reason to believe that three or more particle Green's function
elements are small. Consequently, one should only truncate such
a term if its coefficient is small. For this reason, we have performed
test calculations with the following truncation criterion: 
all terms with more than four $d$ and $d^\dagger$ operators
corresponding to non-partially-filled orbitals are truncated.
In addition, all terms whose coefficient's magnitude is below a cutoff
$\varepsilon$ are eliminated. If there are $N_p$ partially
filled orbitals, then this rule allows terms with up to $4 N_p + 4$ 
$d$'s to appear. In this case we have up to 20 $d$'s, i.e. a
10-particle term.

\begin{table}
\caption{
Same as for Table II, but for a truncation criterion with no
limit on the number of partially occupied terms. Here $N_{3+}$
is the maximum total number of terms involving three or more
particles. Here the O 1s orbital has not been frozen; the 
``exact'' value is taken from a DMRG calculation keeping 750
states: -75.9661.
}\label{tablethree}
\begin{tabular}{l l l}
$\varepsilon$ & $\Delta E$ & $N_{3+}$ \\
\hline
0.001 & 0.0097 & $5.3\times10^5$\\
0.0005 & 0.0025 & $1.4\times10^6$\\
0.0002 & 0.0003 & $3.9\times10^6$\\
\end{tabular}
\end{table}

As shown in Table II,
with this criterion we see substantially better results:
we find that 
in this non-single reference system,
accuracy to fractions of a millihartree is possible.

\section{Flow Equation CD}
It is also possible to formulate CD in terms of a differential
equation. This approach was originally developed independently
by Wegner\cite{wegner} and by Glazek and Wilson,\cite{glazek}
in rather different contexts than we present it here.
We will derive it here as a natural variation of the
Jacobi CD method.
In this approach we introduce a time-like variable
$t$, and the Hamiltonian evolves as $t$ increases. 
First consider a time-dependent Hamiltonian for some fixed
antihermitian operator A:
\begin{eqnarray}
H(t) = e^{tA} H(0) e^{-tA}
\label{Ht}
\end{eqnarray}
Here $H(0)$ is the initial HF Hamiltonian.  We have
\begin{eqnarray}
\frac{dH(t)}{dt} = [A,H(t)]. 
\label{dHt}
\end{eqnarray}
This differential equation form of a CT has long been used in
analytical work, where one integrates $t$ from 0 to some fixed
rotation angle. Here we modify this by making $A$ depend on $H$.
First expand $H$ as follows
\begin{eqnarray}
H(t) = \sum_\alpha a_\alpha(t) h_\alpha .
\label{expandH}
\end{eqnarray}
Each $h_\alpha$ is a product of $d$ and $d^\dagger$ operators, 
and $a_\alpha(t)$ is the corresponding coefficient. Let
\begin{eqnarray}
A(t) = \sum_\alpha s_\alpha a_\alpha(t) h_\alpha .
\label{Asum}
\end{eqnarray}
The $s_\alpha$ are fixed parameters, which we initially consider to
have only three possible values: $\pm 1$, and 0. We set
$s_\alpha$ to
0 if we are not interested in rotating the coefficient of
$h_\alpha$
to 0, because, for example, $h_\alpha$ does not act directly on
the HF state $|0\ket$. For terms we wish to rotate to zero,
we choose the sign of $s_\alpha$ so that (1) $A(t)$ is antihermitian,
and (2) increasing $t$ rotates in the direction to diminish
$a_\alpha(t)$. These conditions are satisfied if $s_\alpha$ is chosen as
the sign of $E_l - E_r$, where $l$ and $r$ are the left and
right states of $h_\alpha$. We evolve $H(t)$ as a sequence of 
infinitesimal CTs, as follows
\begin{eqnarray}
H(t+\delta t) &=& e^{\delta t A(t)} H(t) e^{-\delta t A(t)}
\nonumber \\
&=& H(t) + \delta t [A(t), H(t)] + O(\delta t)^2 .
\label{Htdelta}
\end{eqnarray}
In the limit that $\delta t \to 0$, this is equivalent to
solving the nonlinear differential equation
\begin{eqnarray}
\frac{dH(t)}{dt} = [A(t),H(t)]. 
\label{dHtA}
\end{eqnarray}
Each infinitesimal rotation acts to diminish
each $a_\alpha(t)$ with nonzero $s_\alpha$. 
Since $A(t)$ depends linearly on the
$a_\alpha(t)$,
the rotations become smaller as the $a_\alpha(t)$ decrease. Thus,
we expect the solution of this equation for $t \to \infty$ 
to have $a_\alpha(t) = 0$ if $s_\alpha$ is nonzero. We also expect these
$a_\alpha(t)$ to diminish exponentially with $t$.

If no truncations are made, the solution to this differential
equation for any time $t$ gives an $H(t)$ which is related
to $H$ by an exact CT. This is true also for any choice of the
$s_\alpha$ as long as they satisfy the requirement that if $h_\beta = h_\alpha^\dagger$,
then $s_\beta = -s_\alpha$, ensuring that $A(t)$ is antihermitian.
For numerical efficiency, it is useful to modify the choice
of $s_\alpha$. This is because different terms $h_\alpha$ require
different rotation angles. One would like to make the
exponential decay to zero of each $a_\alpha(t)$ have approximately
the same time constant. If they have widely varying time
constants, the number of steps in integrating the differential
equation will be very large.
We can achieve this by choosing,
for nonzero $s_\alpha$, 
\begin{eqnarray}
s_\alpha = (E_l-E_r)^{-1} .
\label{newsdef}
\end{eqnarray}
Provided $a_\alpha(t) \ll E_l-E_r$, this choice makes the coefficient
of $h_\alpha$ in $A(t)$ the angle $\theta$ required to rotate the
term to zero. This makes the natural time scale for each term
equal to unity. We choose the $s_\alpha$ at the beginning, using
the untransformed HF energies, and never change them; however,
one could also make the $s_\alpha$ depend on $t$.

In Wegner's original flow equation method, rather than the
above forms of $A$ defined in terms of $s_\alpha$, one took
$A = [H^D, H]$, where $H^D$ is the diagonal part of $H$.
This is very similar to the choice $s_\alpha = E_l-E_r$,
assuming all off-diagonal terms are being removed.
However, this choice gives very widely varying time scales,
driving terms with large $E_l-E_r$ to zero much more quickly.
In the sense that the large energy difference terms are removed
first, Wegner's method can be considered a renormalization group
method in itself, and one might stop at some finite time and
study the partially transformed Hamiltonian.
Our choice is much more efficient numerically, assuming one only
wants the $t\to\infty$ limit.

We decribe all the commutator relations in terms of a ``matrix''
$B$
\begin{eqnarray}
[h_\alpha,h_\beta] = \sum_\gamma B^\gamma_{\alpha\beta} h_\gamma .
\label{hcom}
\end{eqnarray}
If a commutator gives a term which is not in the set of
Hamiltonian terms we are keeping, then that term is ignored.
Then the final form for the flow equation CD method
is a set of differential equations
\begin{eqnarray}
\frac{da_\gamma(t)}{dt} = \sum_{\alpha\beta} B^\gamma_{\alpha\beta} 
s_\alpha a_\alpha(t) a_\beta(t)
\label{odeeq}
\end{eqnarray}
which are to be solved numerically. The $B$ matrix was
computed initially and stored in our program. Because of
some regularities in the pattern of nonzero elements of $B$,
the storage could be reduce by a factor of about $N$, the
number of orbitals, from a naive estimate. However, they
could also be recomputed at each step to save storage, at
the expense of computer time. Another approach to save
storage would be to remove a few orbitals at a time. One
could even remove one term at a time by making only one
$s_\alpha$ nonzero, in which case the
flow equation method becomes very similar to the
Jacobi method.
To integrate the coupled differential equations, we use 
a simple fourth order Runge Kutta method with automatic 
step size adjustment. This routine attempted to integrate the differential
equations with an absolute error tolerance of $10^{-8}$, and
we integrated the equations from $t=0$ to $t=20$.

\begin{figure}[tb]
\includegraphics*[height=7cm]{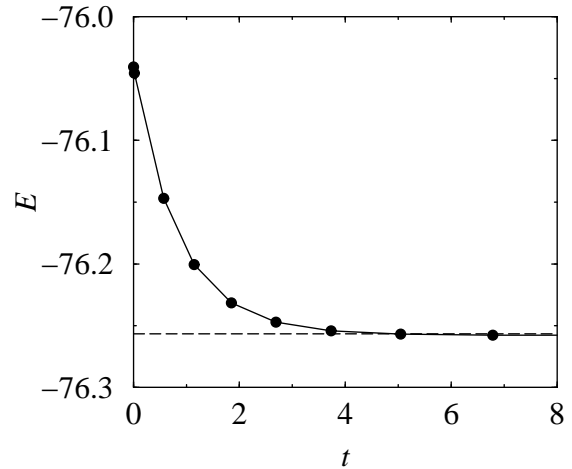}
\caption{
Evolution of the constant term in the renormalized Hamiltonian
as a function of time, for the flow equation CD method.
The system is the same as in Table I.
All two particle terms were retained in H.
The final energy is -76.25795, versus the exact full CI value of 
-76.25662, shown by the dashed line,
for an error of 1.3 milli-hartree.
}
\label{enfigtwo}
\end{figure}

In Figure \ref{enfigtwo}, we show the evolution of the constant term in $H$
as a function of $t$ for the unstretched water molecule.
Only one and two particle terms were retained.
The step sizes used were rather large, and they steadily increased. They
are visible via the circles in the curve. Only twelve steps were
taken, although each RG step in our very crude integrator
required twelve derivative evaluations, Eq. (\ref{odeeq}).
The result for the energy was in error only by about a
milli-hartree.

\begin{figure}[tb]
\includegraphics*[height=7cm]{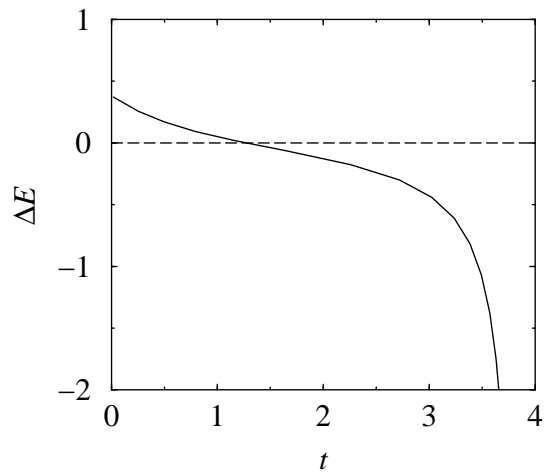}
\caption{
Same as for Fig. \ref{enfigtwo}, but for the stretched water molecule.
Here, the flow CD method retaining only one and two particle
terms is unstable.
}
\label{enfigthree}
\end{figure}

In Figure \ref{enfigthree}, we show similar results for the stretched water
molecule. As in the Jacobi method, CD keeping only two particle
terms is unstable, with the energy tending to $-\infty$. 
We believe that by keeping multiparticle terms one could make
this method perform very well on the stretched water molecule,
just as we found for the Jacobi method.

\section{Removing sets of orbitals}
Another approach for systems such as the stretched water
molecule, which have some strongly correlated orbitals, is
to integrate out many of the non-strongly-correlated-orbitals,
leaving a small but strongly correlated system to solve with CD
retaining many-particle terms, with DMRG, or with another method. 
We first divide the orbitals into
two sets, those to be kept and those to be removed. 
Some of the orbitals to removed will have occupancies near 0,
and some may be core orbitals with occupancies near 2. Due
to the particle-hole transformation, we need make no distinction
between these cases.
Consider the many-particle basis states 
$|s\ket = d^\dagger_i d^\dagger_j \ldots |0\ket$.
Let $|s\ket$ denote all states in which no orbital to be removed
is occupied. Conversely, let $|s'\ket$ denote the rest, in which
at least one orbital to be removed is occupied. We wish to
rotate away all Hamiltonian terms which connect states $|s\ket$
to $|s'\ket$. Let $r$ represent orbitals to be removed. Then the
terms to be removed are described by the following rule:
the terms have one or more $d^\dagger_r$'s, or one or more
$d_r$'s, but do not have both $d^\dagger_r$'s and $d_r$'s.

One finds that typically a few of these terms to be removed, largely by
accident, have $E_r$ and $E_l$ nearly identical, although
neither is close to zero because they include operators
adding or removing high-energy orbitals. To remove
these problem terms requires a large angle of rotation. This can be
disastrous for either the Jacobi or flow equation
method unless many-particle terms are kept. However, some
reflection indicates that these problem terms are likely to be quite
unimportant in terms of their true contribution to the ground
state. Consider an nth order perturbation theory
contribution to the ground state energy. Ignoring the energy
denominators, such a term is proportional to 
\begin{eqnarray}
\bra 0 | h_1 h_2 \ldots h_n | 0 \ket. 
\label{pert}
\end{eqnarray}
Clearly, if this contribution is nonzero, term $h_n$ must have $E_r = 0$,
and term $h_1$ must have $E_l=0$. Thus our problem term with nearly
degenerate nonzero energies cannot contribute in second order. 
For third order,
one might consider $h_2$ to be the problem term. However, for this
term either $|r\ket$ or $|l\ket$ must belong to the set
$|s'\ket$. Removing all the other terms connecting $|s\ket$
to $|s'\ket$ means that either $h_1$ or $h_3$ is to be removed,
since $| 0 \ket$ belongs to $|s\ket$. Thus there is no third
order contribution. The lowest order contribution for such a
term is in fourth order, involving both this term and its
Hermitian conjugate as $h_2$ and $h_3$, or involving two
such terms as $h_2$ and $h_3$. Here $h_4$ takes one from 
$| 0 \ket$ into a higher energy state $|s\ket$, $h_3$ takes
one from $|s\ket$ to $|s'\ket$, $h_2$ then takes one back to
$|s\ket$, and $h_1$ takes one back to $| 0 \ket$. The energy
denominators are well-behaved, since $E_r$ and $E_l$ of the
problem term are not close to zero.

Thus a quite reasonable approach is to rotate away all terms
connecting $|s\ket$ to $|s'\ket$ except those whose energy
difference $|E_r-E_l|$ is below some cutoff $d$. The terms below
the cutoff are retained during the CD process, during which
time they may change due to other terms being rotated away.
After the CD is complete, one then can discard all terms
having any $d_r$ or $d^\dagger_r$ operators, which will include
these problem terms. One can also solve the CD-transformed $H$
before truncation, using another method, and check that the
occupancies of the removed orbitals are very close to zero.

\begin{table}
\caption{
Results for the flow equation method applied to
integrate out a set of orbitals, coupled with DMRG to
solve the resulting Hamiltonian. The system is the stretched
water molecule of Table II, with 25 orbitals. The first column
tells how many orbitals were removed, specified as having
the highest single particle energies in the
particle-hole-transformed Hamiltonian.
The parameter $d$ is the
lower limit on the energy difference of an operator for it
to be removed. $\Delta E_{\rm CD}$ is the error in the energy,
as computed by DMRG,
relative to the full CI energy (-75.95227) after CD has been
performed to eliminate orbitals, but with all 25 orbitals still present. 
$n_{\rm max}$ is the largest occupancy of any of the orbitals
which have been ``rotated away''. $\Delta E_{\rm CDT}$ is
the error in the energy, computed by DMRG, after CD and after
truncation of the rotated orbitals. The $^*$ indicates that the
ground state in this diagonalization has a clearly erroneous
orbital occupancy pattern, indicating that it is a low lying excited state
which has dropped below the true ground state. The true ground
state occupancy pattern reappeared upon truncation of the
rotated orbitals.
}\label{tablefour}
\begin{tabular}{c l l l l}
Orbitals & $d$\qquad\qquad & $\Delta E_{\rm CD}$ & $n_{\rm max}$ & $\Delta E_{\rm CDT}$ \\
Removed&& &&\\
\hline
8 & 0.5 & -0.0003 & $1\times10^{-5}$ & -0.0003\\
13 & 0.5 & -0.0003 & $9\times10^{-6}$ & -0.0003\\
17 & 0.5 & 0.016 & $6\times10^{-5}$ & 0.016\\
17 & 1.0 & 0.011 & $3\times10^{-3}$ & 0.018\\
20 & 0.5 & 0.007$^*$ & $3\times10^{-6}$ & 0.014\\
21 & 0.5 & 0.011 & 0.01 & 0.012\\
\end{tabular}
\end{table}

In Table IV we show the results of such calculations for
the stretched water molecule. There are three sources of error
in these calculations. First is the DMRG error, 
typically near $0.0002$ mH, keeping 400-600 states, which is
small enough to show the other sources of error.
Second is the error from performing CD keeping only one and two particle
terms. This is given by $\Delta E_{\rm CD}$. Increasing $d$, or
removing fewer orbitals improves $\Delta E_{\rm CD}$. Third,
there is the energy from throwing away the removed orbitals
after CD. This is measured by the difference between $\Delta
E_{\rm CD}$ and $\Delta E_{\rm CDT}$, and also by the maximum
occupancy of the removed orbitals $n_{\rm max}$.
We find that $d$ can be made quite large: $0.5$ is always fine,
whereas $1.0$ can be too large. We also find that we can
remove up to about one half of the orbitals and incur only
a very small error, even only keeping one and two particle
terms. For the resulting small system even full CI would be
a very easy calculation.
Even removing all but four of the orbitals we get a
reasonable result. We have not carried out any similar
calculations keeping many-particle terms, but we can deduce
the probable outcome. Since all the rotation angles $\theta$ are rather
small in this procedure, four particle terms, which can come in
only as $\theta^2$, would be neglible. Three particle terms come
in as $\theta$, and if such a term only involved the retained
orbitals it presumably would have both $E_l$ and $E_r$ small and
it could give a substantial contribution to the energy of order
$\theta$. Three particle terms involving removed orbitals would 
have $E_l$ or $E_r$ reasonably large, and would only 
contribute to the energy via second order perturbation terms, thus
coming in as $\theta^2$, which could be neglected.
In short, we expect that keeping three particle terms involving the
retained orbitals only would be a very accurate
approach for removing more than half of the orbitals.

We would like to conclude this section with an argument that
the proper way to separate the treatment of high-energy from
low-energy orbitals is by using an effective Hamiltonian to
remove the high energy orbitals, as we have done, rather
than any wavefunction based approach. We will make this argument
via a trivial $3\times3$ matrix, designed to have some of the
crucial features of a strongly correlated/multireference system.
Define the matrix
\begin{eqnarray}
H(\varepsilon,\delta) = 
\left( \begin{array}{ccc}
0 &  \varepsilon & \delta \\
\varepsilon &  \varepsilon & 1 \\
\delta &  1 & 10
\end{array} \right).
\label{hmat}
\end{eqnarray}
The third row and column represent a high energy orbital, which
we would like to treat separately from the first two nearly
degenerate rows and columns. We will consider the parameter
values ($\varepsilon=0.1$, $\delta = 1$), 
($\varepsilon=0.1$, $\delta = 0.5$),  and
($\varepsilon=0$, $\delta = 1$). For these three parameters
we find the following ground state energies and eigenvectors
(respectively):
-0.099, and (0.995,-0.0098,-0.098); 
-0.064, and (0.789,-0.614,-0.022); and
-0.196, and (0.700,0.700,-0.137).
Now suppose we wanted to solve this system in two steps, first
treating the third ``orbital'', then next the other two, using a
wavefunction approach. In treating the third orbital we insist
that we ignore the small parameter $\varepsilon$; otherwise we
are treating the whole matrix together. We imagine that
we have some perturbative method for obtaining the third
component of the wavefunction, ignoring $\varepsilon$; 
with this fixed, then we obtain the first two components, 
taking $\varepsilon$ into account. However, comparing the
first and third sets of parameters, we see that the
third component $\psi_3$ depends strongly on $\varepsilon$, so
this method must fail.

Alternatively, we might imagine first treating the first two
rows and columns separately, ignoring $\delta$ and
finding the ratio of components
$\psi_1/\psi_2$, and then subsequently using $\delta$ to
fix $\psi_3$. In this case, comparing the first and second
parameter sets, we see that $\psi_1/\psi_2$ depends strongly
on $\delta$, so that this method fails. In short, to treat this
problem successfully, wavefunction based approaches must treat both 
$\delta$ and $\varepsilon$ simultaneously.

Now consider a simple CT approach. Rather than using the Jacobi or
flow equation method, we use a less sophisticated,
but well-known perturbative CT method for removing the third
row and column.\cite{lax}  In this case, we find that the
second-order change
in the upper left $2\times2$ portion of the matrix, due
to the third row and column, is
\begin{eqnarray}
\Delta H_{ij} = H_{i3} H_{3j} \frac{1}{2} (\frac{1}{E_i-E_3} + 
\frac{1}{E_j-E_3})
\label{DelH}
\end{eqnarray}
where $E_i = H_{ii}$. (The general formula is obtained by
replacing 3 by $k$ and summing over all orbitals to be removed
$k$.) $\varepsilon$ appears only in the energy
denominators, as a small correction; we ignore it by setting it 
to zero there. We
obtain $H^{\rm eff} = H + \Delta H$ as
\begin{eqnarray}
H^{\rm eff}(\varepsilon,\delta) = 
\left( \begin{array}{cc}
-\frac{\delta^2}{10} &  \varepsilon - \frac{\delta}{10} \\
\varepsilon - \frac{\delta}{10} & \varepsilon - \frac{1}{10}
\end{array} \right).
\label{heff}
\end{eqnarray}
The ground state energies and eigenvalues for $H^{\rm eff}$ for
the three cases are 
-0.1, and (1.0,0.0); 
-0.064, and (0.788,-0.615); and
-0.2, and (0.707,0.707). These results compare very nicely to
the exact results for the full matrix. Indeed, they must; the
procedure is well controlled, with large energy denominators.

In order to properly separate the two parts of the problem in
a wavefunction-based approach, one needs to allow a set of
possible wavefunctions to represent the high energy states, rather
than a single part of a wavefunction. Such an approach is
embodied in the DMRG method, which chooses the optimal set
of states to represent each part of the system.

\section{Discussion}
CD is size-consistent: if one duplicated the Hamiltonian for
a system, corresponding to having two molecules separated by
a large distance, and put in no interaction terms between the 
two systems, then no interaction terms would ever be
generated and each system would behave identically under the CTs.
The energy would be double the energy for one system.

The calculation time for CD generally scales identically 
with the number of orbitals $N$ for the Jacobi and
flow equation methods. Consider first the methods which
directly determine the ground state, rather than removing orbitals
first. There are of order
$N_{\rm occ}^2 N_{\rm unocc}^2$ terms $r$ which connect directly
to $|0\ket$, where $N_{\rm occ}$ ($N_{\rm unocc}^2$) are the number
of occupied (unoccupied) orbitals, which one needs to remove. 
Not all other terms $s$ connect to any term $r$; if one is discarding
all three particle terms, then there must be two orbital indices
matching in $r$ and $s$ to get a contribution. 
Thus each term $r$ connects to of order $N^2$ terms $s$. Hence
the total calculation time scales as $N_{\rm occ}^2 N_{\rm
unocc}^2 N^2$, or roughly $N_{\rm occ}^2 N^4$ or more roughly $N^6$.
This is comparable to a singles and doubles CI or coupled cluster.
If one treats only the terms $r$ with large angles, using
second order perturbation theory for the rest, the calculation
time would be reduced but the scaling is more difficult to
analyze. However, from the results of Fig. 1 it is tempting
to estimate the number of terms needed to be rotated as about
$N^2$, leading to an overall scaling of $N^4$ (plus a time
of order $N^5$ for the initial HF change of basis). Of course,
studies of systems of various sizes are necessary to determine
the true dependence on $N$. (It is also challenging to write
efficient programs for CD which exploit the potentially
favorable scaling: if one is not careful, one may find one's
program spending most if its time performing commutators very
slowly for terms with small coefficients which are later
discarded.)
One could also only rotate the
largest $N^2$ terms using the flow equation method,
and then use perturbation theory for the rest of the terms,
leading to similar scaling with system size.
There are also other
variations of CD with good scaling. Note that if one
does CD but restricts the terms $s$ to be either $r^\dagger$, or
a diagonal term whose indices all match those in $r$, then
one obtains an $o(N^4)$ method closely related to second 
order perturbation theory. A presumably more accurate
$o(N^5)$ method is obtained if
one restricts $s$ so that three indices must match those
in $r$, rather than two. 
For CD where one removes sets of orbitals,
keeping one and two particle terms, the scaling to remove each
orbital is $o(N_{\rm occ}^2 N^3)$, for a total of 
$N_{\rm occ}^2 N^4$ to remove a finite fraction of the orbitals.

Let us discuss in more detail how to think about the canonical
transformations\cite{wagner}. Thus far, we have taken the view that we apply
a CT to get a new Hamiltonian
\begin{eqnarray}
\tilde H = e^A H e^{-A} ,
\label{Htil}
\end{eqnarray}
which has different coefficients from $H$,
but is written in terms of the same operators
\begin{eqnarray}
\tilde H = \sum_\alpha \tilde a_\alpha h_\alpha .
\label{tilHtila}
\end{eqnarray}
The new Hamiltonian has the same eigenvalues as the old, and one
can reconstruct the eigenvectors: if
\begin{eqnarray}
\tilde H |\psi\ket = E |\psi\ket,
\end{eqnarray}
then
\begin{eqnarray}
H e^{-A} |\psi\ket = E e^{-A} |\psi\ket,
\end{eqnarray}
so that $e^{-A} |\psi\ket$ is the corresponding eigenvector of
$H$.  
One could also define new operators $d_i$ and $d^\dagger_i$ as
\begin{eqnarray}
\tilde d_i = e^A d_i e^{-A} ,
\label{tild}
\end{eqnarray}
where the same expression applies for $d^\dagger_i$. Since
\begin{eqnarray}
e^A d_i d_j e^{-A} = 
e^A d_i e^{-A}e^A d_j e^{-A} = \tilde d_i \tilde d_j,
\label{etild}
\end{eqnarray}
one could equally well write $\tilde H$ as
\begin{eqnarray}
\tilde H = \sum_\alpha a_\alpha \tilde h_\alpha .
\label{tilHtilh}
\end{eqnarray}
Here $\tilde h_\alpha$ is a product of $\tilde d_i$ operators
with the same orbital indices and order as $h_\alpha$.

This form, Eq. (\ref{tilHtilh}), 
is not especially useful, since the coefficients of the
Hamiltonian are not any more diagonal than in $H$. 
A more useful expression comes from writing
\begin{eqnarray}
H = e^{-A} e^A H e^{-A} e^A = e^{-A} \tilde H e^A.
\label{HtilAA}
\end{eqnarray}
If we define new operators $\bar d$ using the inverse CT,
\begin{eqnarray}
\bar d_i = e^{-A} d_i e^A ,
\label{bard}
\end{eqnarray}
then
\begin{eqnarray}
H = \sum_\alpha \tilde a_\alpha \bar h_\alpha ,
\label{tilHbarh}
\end{eqnarray}
where $\bar h_\alpha$ is defined analogously to $\tilde h_\alpha$.
We see that in terms of the $\bar d$ operators, the original
Hamiltonian has the more diagonal form for the coefficients of 
$\tilde H$. This means that one should think of 
$\bar d^\dagger_i$ as the operator which creates a
quasiparticle, not $\tilde d^\dagger_i$. In particular, suppose
$A$ fully diagonalizes $H$, in which case any Slater determinant
is an eigenstate of $\tilde H$. For any orbital $i$
\begin{eqnarray}
\tilde H d^\dagger_i |0\ket  = \varepsilon_i d^\dagger_i |0\ket,
\label{epsdi}
\end{eqnarray}
from which we obtain
\begin{eqnarray}
H \bar d^\dagger_i e^{-A} |0\ket  = \varepsilon_i 
  \bar d^\dagger_i e^{-A} |0\ket . 
\label{bardi}
\end{eqnarray}
We see that $\bar d^\dagger_i$ creates a new exact eigenstate
from the ground state $e^{-A} |0\ket$, 
containing an extra particle associated
with orbital $i$. This defines $\bar d^\dagger_i$ to be
a quasiparticle creation operator. It creates a ``dressed''
electron, with correlations built in. Because of the
correlations built in,
three and more particle terms can appear in $\tilde H$.
Note that if one has exactly
diagonalized $H$ with $A$, then one can create {\it all} of the
excited states by successively applying $\bar d^\dagger_i$'s to
$e^{-A} |0\ket$. 

The formulation of CD in terms of exponentials of operators
has much in common with coupled cluster methods (CC). In coupled
cluster methods, the ground state wavefunction is written as $e^T |0\ket$.
Usually $T$ is not antihermitian, but in some less common versions of CC,
it is, and usually the CC equations are derived using (formally)
a similarity transformation of H.\cite{cc} 
One difference between the two is that in CD
we never explicitly write down $A$; rather, we perform 
a sequence of transformations $A_1, A_2,\ldots A_n$, which implicitly
define the complete transformation $e^A = e^{A_n} \ldots e^{A_1}$.
(In the flow equation method this sequence is
continuous.)
Based on the similar expressions for the ground state, one
might expect CD and CC to have similar errors, and our results
are generally consistent with this. However, the overall point
of view between CD and CC is fundamentally different: CC is
approximating the ground state, whereas CD is progressively transforming
the Hamiltonian into a diagonal form. 
The point of view of CD makes certain
approaches natural and manageable, including removing sets of
orbitals, extracting excited states, and utilizing
renormalization group ideas.

Furthermore, CD, in its various approximate forms, makes its
truncations of $H$ at each transformation. These intermediate
truncations make tractable the use of unitary transformations,
rather than non-unitary similarity transformations. Such
continuous truncations are familiar from RG methods in
statistical mechanics. One way of understanding their usefulness
is to consider diagonalizing a matrix with an approximate
second order unitary transformation, as in the previous section.
Here, however, we consider transforming the whole matrix this way.  Except
for matrices which are nearly diagonal to start with, this
second order approach would work very poorly. However, if one
makes a sequence of second-order unitary transformations, 
each having very small
rotation angles, the method becomes accurate; in fact, it is
exact in the continuous limit. This is analogous to integrating
an ordinary differential equation very precisely with a sequence 
of very small time steps, using a low order integration method.
This is also why the flow equation CD method, without truncation,
is exact even though only a first order commutator appears
in the equation.
The truncation of many particle terms is not really 
analogous to throwing away higher order commutators, and so
CD with truncation is not exact. However, there is no reason
{\it a priori} to expect that CD, with its continous truncations, 
should be worse than CC.

Now let us briefly mention how to obtain
excited states. Suppose one wants to know the energy of an
excited state which has a large overlap with the state
$d^\dagger_i |0\ket$. One needs to remove all off-diagonal
terms which do not destroy this state, such as 
$d^\dagger_j d^\dagger_k d^\dagger_l  d_i$, plus their Hermitian
conjugates. This includes terms such as 
$d^\dagger_j d^\dagger_k d^\dagger_l d^\dagger_m$, which one
would already remove to get the ground state. It may happen
that some of these new terms to remove would require large rotation angles,
in which case one might want to remove most of the orbitals
first. Note that if one removes a large number of orbitals,
a full diagonalization obtaining all excited states of the
remaining orbitals may be quite manageable. One might also try
to remove {\it all} off-diagonal terms in $H$, in which case
{\it all} the excited state energies could be obtained by inspection!
Note that the work for removing all off-diagonal terms in $H$ would
still scale as $N^6$. However, in this case, there would be many
terms with nearly degenerate $E_r$ and $E_l$ which would cause
problems. We leave exploration of these approaches for future
work.

Let us also briefly mention calculation of expectation values of
operators in the ground state, $\bra A \ket$. One approach
is simply to apply the same CTs to $A$ as one has applied to
$H$, truncating many-particle terms in a similar fashion, 
to get $\tilde A$, and then evaluate $\bra 0 | \tilde A |0 \ket$.
Another approach would be to obtain an approximate expression
for the ground state $| \psi \ket$ in the original HF basis,
by applying $\exp(A)$ successively to $|0\ket$ for each CT in reverse
order, again with some truncation rules. Again, we leave
exploration of these approaches for future work.

\section{Conclusions}
We have outlined a numerical approach, canonical
diagonalization, for treating a variety of quantum many
body problems. CD is quite different from most existing
methods for treating such problems: it does not utilize 
approximate wavefunctions, semiclassical approximations,
path integrals, perturbation theory, 
or Monte Carlo. Instead, the second quantized Hamiltonian
is transformed directly, using canonical transformations,
to put it into a diagonal form.

We have demonstrated CD on {\it ab initio} quantum chemical
calculations for a small molecule. CD appears to be quite
competitive with the best alternative quantum chemical methods,
such as the coupled cluster method, even in this early stage
of its development. Unlike many other approaches, CD can be
used to treat systems where the ground state has a small overlap
with the Hartree Fock state. It can also be used to remove high
energy orbitals from the problem, leaving a smaller problem
which can be treated with other methods, such as DMRG.
Although we have not yet tested the ability of CD to obtain
excited states, there is no fundamental difference between
the ground state and an excited state within CD, and we have
outlined specific methods to obtain excited states.

One of the principle future uses of
CD could be to derive simple model Hamiltonians, much studied
in condensed matter physics, directly
from {\it ab initio} calculations. Currently, deriving model Hamiltonians
is an art which involves educated guesses for the proper
model forms coupled with the matching of completely separate
solutions for the {\it ab initio} and model systems.
CD may be able to unify this approach into a controlled
single procedure.

\begin{acknowledgments}

We would like to acknowledge valuable discussions Alex Maradudin,
F. Wegner, and A. Mielke. We particularly thank
Garnet Kin-Lic Chan for a numerous comments and discussions.
We acknowledge the support of the NSF under grant
DMR98-70930.

\end{acknowledgments}
{}

\end{document}